\documentstyle[psfig]{mn2e}

\newif\ifAMStwofonts
\AMStwofontstrue

\title[Red Clump population corrections]
         {An empirical test of the theoretical population corrections to the
Red Clump absolute magnitude}

\author[Susan M. Percival \& Maurizio Salaris]
       {Susan M. Percival$^{1}$ and Maurizio Salaris$^{1}$ \\
$^1$Astrophysics Research Institute, Liverpool John Moores
        University, Twelve Quays House, Egerton Wharf, Birkenhead CH41 1LD, 
	UK }
\date{Accepted 2003 ???.
      Received 2003 ???;
      in original form 2003 ???}

\pagerange{\pageref{firstpage}--\pageref{lastpage}}
\pubyear{2003}

\begin{document}

\maketitle

\label{firstpage}

\begin{abstract}

The mean absolute magnitude of the local red clump (RC), $M^{RC}_{\lambda}$, 
is a very well determined quantity due to the availability of accurate 
$Hipparcos$ parallaxes for several hundred RC stars, potentially allowing 
it to be used as an accurate extra-galactic distance indicator.  
Theoretical models predict that the RC mean magnitude is dependent on 
both age and metallicity and, furthermore, that these dependencies are 
non-linear.  This suggests that a population correction, 
$\Delta M_{\lambda}^{RC}$, based on the star formation rate (SFR) and 
age-metallicity relation (AMR) of the system in question, should be applied 
to the local RC magnitude before it can be compared to the RC in any other 
system in order to make a meaningful distance determination.  

Using a sample of 8 Galactic open clusters and the Galactic Globular Cluster 
47~Tuc, we determine the cluster distances, and hence the RC absolute 
magnitude in $V$, $I$ and $K$, by applying our empirical main sequence 
fitting method, which utilizes a large sample of local field dwarfs 
with accurate $Hipparcos$ parallaxes.  
The 9 clusters have metallicities in the range
$-0.7 \leq {\rm [Fe/H]} \leq +0.02$ and ages from 1 to 11 Gyr, enabling
us to make a quantitative assessment of the age and metallicity dependences 
of $\Delta M_{\lambda}^{RC}$ predicted by the theoretical models of Girardi
\& Salaris (2001) and Salaris \& Girardi (2002).
We find excellent agreement between the empirical data and the models in 
all 3 pass-bands, with no statistically significant trends or offsets, thus 
fully confirming the applicability of the models to single-age, 
single-metallicity stellar populations.  

Since, from the models, $\Delta M_{\lambda}^{RC}$ is a complicated function
of both metallicity and age, if this method is used to derive distances to  
composite populations, it is essential to have an accurate assessment of the 
SFR and AMR of the system in question, if errors of several tenths of a 
magnitude are to be avoided.
Using recent determinations of the SFR and AMR for 4 systems -- the LMC, SMC, 
Carina and the solar neighbourhood -- we examine the quantity
${I}^{RC}_{\rm obs} - {K}^{RC}_{\rm obs}$, which is the difference between 
the mean magnitude of the RC in the $I$-band and the $K$-band.  Comparing the 
theoretical predictions with the most recent observational data, we find
complete agreement between the observations and the models, thus confirming
even further the applicability of the population corrections predicted from 
theory.

\end{abstract}


\section{Introduction}
\label{intro}
In recent years much work has been devoted to studying the suitability of Red 
Clump (RC) stars as a distance indicator.  These core helium-burning stars
are essentially in an identical evolutionary phase to those which make up the 
horizontal branch in globular clusters.  However, in intermediate and higher 
metallicity systems only the red end of the distribution is seen, forming a
clump of stars in the colour-magniutude diagram - hence the name.
In an infrared study of metal-rich globular clusters,
Kuchinski et al. (1995) realised that the absolute $K$-band magnitude of the 
horizontal branch (effectively, the RC) displayed very little 
cluster-to-cluster
variation, and therefore suggested it could potentially be used as a standard 
candle for distance determinations.  However, at that time, no absolute
calibration of the RC magnitude was available and so distances could only be 
measured differentially, with respect to a specific cluster. 
Interest in the RC as a potential extra-galactic distance indicator was 
rekindled a few years later due to the availability of accurate
$Hipparcos$ (ESA 1997) parallaxes (and hence absolute magnitudes) for several 
hundred local RC stars, enabling a very precise calibration of their mean 
absolute magnitude.  In fact, RC stars are currently the only extra-galactic 
distance indicator which can be tied precisely to the $Hipparcos$ distance 
scale, due to the lack of a significant number of RR Lyraes with accurate 
parallax determinations (see e.g. Fernley et al. 1998) and the current 
uncertainties in the metallicity dependence of the Cepheid period--luminosity 
relation (Feast 2003, and references therein).

Paczy\'nski \& Stanek (1998) made the first attempt to provide a precise 
measurement of the local RC absolute magnitude from $Hipparcos$ data by
fitting the $I$-band magnitudes of RC stars to a Gaussian function, 
superimposed on a background distribution of red giant branch (RGB) stars, to 
find a peak value of $M^{RC}_{I} = -0.28$.
This early work focussed on the $I$-band since, observationally, the $I$-band 
magnitudes of RC stars appeared to show no dependence on colour, unlike those 
in the $V$-band, hence it was thought that the $I$-band magnitude of RC stars
may provide a truly `standard' candle.  The initial value from Paczy\'nski 
\& Stanek (1998) was subsequently revised slightly to $M^{RC}_{I} = -0.23$ by 
Stanek \& Garnavich (1998).  This was then 
used by Stanek, Zaritsky \& Harris (1998) to make the first direct comparison 
with the RC in the LMC, yielding a dereddened distance modulus of
$\mu_{0,LMC} = 18.065 \pm 0.031_{\rm statistical} \pm 0.09_{\rm systematic}$.
Contemporaneously, an almost identical result of
$\mu_{0,LMC} = 18.08 \pm 0.03_{\rm statistical} \pm 0.12_{\rm systematic}$
was found by Udalski et al. (1998) using the same method.  The implied 
LMC distance from these
results is short, even by the standards of the `short' distance scale, and 
they are in significant disagreement with the LMC distance modulus of 
$\mu_{0,LMC} = 18.50 \pm 0.10$ adopted by the $Hubble~Space~Telescope$ 
Extragalactic Distance Scale Key Project (Freedman et al. 1994).

It was soon pointed out that models of core helium-burning stars predict that 
the RC luminosity is dependent on both age and metallicity, causing 
differences of up to 0.6 mag in the mean value of $M^{RC}_{I}$ between 
different stellar populations (Cole 1998; Girardi et al. 1998).  
The implication therefore, is that any distance derivation to an extragalactic
system should first account for population differences between the local RC 
and the RC of the system in question, for the results to be meaningful.
In an attempt to address this problem, Udalski (2000) examined the
metallicity dependence of $M^{RC}_{I}$ in the solar neighbourhood from the 
empirical data.  He found a weak dependence of $M^{RC}_{I}$ on [Fe/H]
at the level of $\sim$ 0.13 mag ${\rm dex}^{-1}$, and consequently derived an
$I$-band correction of $-$0.07 mag for the LMC.
Subsequent work has focussed on tightening the constraints on the metallicity
dependence of $M^{RC}$ in several pass-bands, whilst at the same time, 
investigating the effects of age.

Based on the models of Girardi et al. (2000), predicted mean RC magnitudes 
for a large range of ages (0.5$-$12 Gyr) and metallicities 
($-1.7 \leq {\rm [Fe/H]} \leq +0.2$) are given by Girardi \& Salaris (2001,
hereafter GS01, $V$ and $I$-band) and Salaris \& Girardi (2002, hereafter 
SG02, $K$-band).  In order to test these
predictions the effects of age and metallicity must be examined separately.
To this end, a number of recent studies have examined small samples of Galactic
open clusters, each of single age and chemical composition, enabling a 
comparison of the observed RC magnitudes with those predicted from theory 
(Sarajedini 1999; Twarog, Anthony-Twarog \& Bricker 1999; GS01; 
Grocholski \& Sarajedini 2002).  Qualitatively, these studies find
broad agreement with the models in all 3 pass-bands ($V$, $I$ and $K$),
confirming that `population corrections' to the RC absolute magnitude are
necessary before it can be used as an accurate extragalactic distance 
indicator.  It is important to note here that the metallicity dependence
found by Udalski (2000) is derived from local stars in the disc and cannot
be regarded as universally applicable since, if we accept the need for 
population corrections at whatever level, the star formation history and 
age-metallicity relation are likely to be key factors in determining the 
metallicity dependence of the RC magnitude.  In fact, GS01 model the RC in 
the solar neighbourhood and reproduce very well the metallicity dependence 
found by Udalski, giving further weight to the applicability of the models. 

In this work, we give a more quantitative assessment of the accuracy of the
theoretical predictions for the RC population corrections, over a range of 
ages and metallicities.  
We also aim to improve on previous work in several ways.  Firstly, 
determination of the RC absolute magnitude for a particular cluster requires 
knowledge of the cluster distance.  All previous studies involving samples of 
galactic open clusters have used distances which have been derived using
methods based on fitting the cluster sequences to theoretical isochrones.  For 
example, Sarajedini (1999) employs cluster distances which are derived 
relative to that of M67, the distance to M67 having been determined by fitting
to theoretical isochrones.  These methods can potentially introduce 
systematic errors which depend on the particular choice of isochrones.  
In this work, we employ the 
purely empirical MS-fitting method, based on a large sample of local field 
dwarfs with accurate $Hipparcos$ parallaxes, developed in Percival, Salaris 
\& Kilkenny (2003, hereafter Paper I), thus removing a potentially large 
source of systematic error.  Furthermore, we ensure complete consistency in 
the analysis by studying $V$-, $I$- and $K$-band data simultaneously, 
utilizing the same distance modulus, age estimate and reddening for each 
cluster in all 3 bands.

Secondly, using Monte Carlo methods, we quantify the effects of sampling on 
the luminosity of the RC in clusters of different ages
and metallicities, in terms of its (empirically) measured magnitude and the
likely error.  The same simulations also allow us to quantify any 
offsets between mode, median and mean values for the RC absolute magnitude,
$M^{RC}_{\lambda}$.  It should be noted that the model values tabulated in 
GS01 and SG02 are the predicted mean values for $M^{RC}_{\lambda}$ whilst, 
from empirical data, generally the median of the distribution is 
measured.
These values are expected to differ since the distribution of magnitudes
in the RC is not symmetric -- the mode of the magnitude distribution is 
centred on the zero age horizontal branch, and the distribution has a tail 
towards brighter magnitudes as the RC stars evolve to higher luminosities.

A third important element in this work, particularly when considering the 
population effects in the LMC, is the inclusion of 4 clusters with ages 
$\leq$ 2 Gyr.  These young clusters are important since, for populations with 
recent star formation (like the LMC), a large fraction of RC stars have ages 
in the range 1$-$2 Gyr.  The models predict large, and rapid, changes in 
$M^{RC}$ in all 3 pass-bands for stars with ages $<$ 2 Gyr and so it is
vitally important to test these predictions empirically, and quantitatively 
assess their validity.   

The layout of the paper is as follows:
In Section~\ref{RCasDI} we review the basic method of using the RC as a 
distance indicator, in Sect.~\ref{clsample} we give details of the clusters
used in this study and in Sect.~\ref{method} we describe the methods 
involved in deriving cluster distances, and determining the RC
magnitude, both empirically and from the theoretical models.
Sect.~\ref{results} presents our results and in Sect.~\ref{disc} we discuss 
some implications of these results and make our conclusions.


\section{The RC as a Distance Indicator -- Basic Method}
\label{RCasDI}

In order to utilize the Red Clump as an extragalactic distance indicator, once
the apparent magnitude of the RC, $m_{\lambda}^{RC}$, is measured, the 
distance modulus, $\mu_{0} = (m-M)_{0}$, to a particular galaxy is given by:
\begin{equation}
\mu_{0,galaxy} = m_{\lambda}^{RC} - M_{\lambda}^{RC} - A_{\lambda} + \Delta M_{\lambda}^{RC}
\label{dm_eq1}
\end{equation}
where $m_{\lambda}^{RC}$ is the apparent magnitude of the RC in a particular
pass-band, $M_{\lambda}^{RC}$ is the `zero-point' absolute RC magnitude
given by the local RC stars, $A_{\lambda}$ is the extinction, and 
$\Delta M_{\lambda}^{RC}$ is the population correction to the RC magnitude.

The most recent determinations of the local RC absolute magnitude,
$M_{\lambda}^{RC}$ are given by Alves et al. (2002), and are $0.73\pm0.03$, 
$-0.26\pm0.03$ and $-1.6\pm0.03$ mag in, respectively, the $V$, $I$ and $K$ 
pass-bands. 
$\Delta M_{\lambda}^{RC}$, the population correction, is defined as the 
difference between the local RC absolute magnitude and that of the RC of the 
system in question i.e.
$\Delta M_{\lambda}^{RC} \equiv M_{\lambda}^{RC}(local) - M_{\lambda}^{RC}(galaxy)$.
As we have already seen, theory predicts that $\Delta M_{\lambda}^{RC}$ is 
dependent on both age and metallicity.

To test for and quantify these dependencies, we need to employ a sample of
galactic open clusters i.e. simple populations, each of single age and single 
metallicity, so that the effects of age and metallicity can be disentangled.
From eq.~\ref{dm_eq1}, it can be seen that the population correction, 
$\Delta M_{\lambda}^{RC}$, for a particular cluster will be given by:
\begin{equation}
\Delta M_{\lambda}^{RC} = \mu_{0,cluster} - m_{\lambda}^{RC} +  M_{\lambda}^{RC} + A_{\lambda}
\label{dm_eq2}
\end{equation}
In this equation, $m_{\lambda}^{RC}$, $M_{\lambda}^{RC}$ and $A_{\lambda}$
are known quantities from the literature (see sect.~\ref{clsample} for 
full details) and the distance moduli to the individual clusters, 
$\mu_{0,cluster}$ are to be determined.

As already noted, all previous studies have relied on cluster distances 
derived from MS fitting to theoretical isochrones (i.e. using stellar models),
which potentially introduces systematic errors.  
The aim of this work is to quantify the population corrections via a 
completely model independent method by utilizing our purely empirical 
MS-fitting procedure (Paper I) to derive the cluster distances.  
This method employs a sample of 54 local unevolved field dwarfs with 
accurate $Hipparcos$ parallaxes (errors are generally $< 5\%$) for which we 
acquired new $BVI_{C}$ photometry.  From fitting to this sample in both the 
$(B-V)$ and $(V-I)$ colour planes we derive a Hyades distance modulus of 
$(m-M)_{0}=3.33 \pm 0.06$ -- exactly reproducing the distance directly 
measured from parallaxes of the Hyades stars (Perryman et al. 1998).
Hence we are able to tie the absolute magnitude of the RC in each of the open 
clusters in our sample to the $Hipparcos$ distance scale.  Since the absolute
magnitude of the local RC is also determined from $Hipparcos$ parallaxes, 
systematic errors are minimized and we can provide a completely 
self-consistent assessment of the age and metallicity dependencies for the 
RC population corrections.

\section{Choosing the Cluster Sample}
\label{clsample}

\begin{table*}
\caption{Cluster data --  metallicities from Gratton (2000), reddenings from 
Twarog et al. (1997).  
Cluster ages, dereddened distance moduli and absolute RC magnitudes in $V$, 
$I$ and $K$ are those derived in this work.}
\label{tab_clus}
\begin{tabular}{lrrrrccc} \hline
Cluster & \multicolumn{1}{c}{[Fe/H]} & \multicolumn{1}{c}{$E(B-V)$} & 
\multicolumn{1}{c}{Age in Gyr} & \multicolumn{1}{c}{$(m-M)_{0}$} & 
$M^{RC}_{V}$ & $M^{RC}_{I}$ & $M^{RC}_{K}$\\
\hline
M67      & $+0.02\pm0.06$ & $0.04\pm0.02$ &  $4.0\pm0.5$ & $ 9.60\pm0.09$ & $0.820\pm0.115$ & $-0.203\pm0.103$ & $-1.614\pm0.097$\\
NGC 2477 & $ 0.00\pm0.08$ & $0.23\pm0.02$ &  $1.0\pm0.3$ & $10.74\pm0.08$ & $0.915\pm0.115$ & $-0.030\pm0.102$ & $-1.304\pm0.095$\\
NGC 188  & $-0.03\pm0.06$ & $0.09\pm0.02$ &  $6.0\pm0.5$ & $11.17\pm0.08$ & $0.988\pm0.106$ & $-0.091\pm0.092$ &  *****\\
NGC 7789 & $-0.13\pm0.08$ & $0.29\pm0.02$ &  $1.6\pm0.3$ & $11.22\pm0.07$ & $0.800\pm0.100$ & $-0.140\pm0.085$ &  *****\\
Be 39    & $-0.15\pm0.09$ & $0.11\pm0.02$ &  $7.5\pm1.0$ & $12.97\pm0.09$ & $0.954\pm0.112$ & $-0.107\pm0.100$ & $-1.434\pm0.094$\\
NGC 2204 & $-0.38\pm0.08$ & $0.08\pm0.02$ &  $1.7\pm0.3$ & $13.12\pm0.08$ & $0.481\pm0.113$ & $-0.435\pm0.100$ & $-1.676\pm0.093$\\
Mel 66   & $-0.38\pm0.06$ & $0.14\pm0.02$ &  $4.5\pm0.5$ & $13.22\pm0.07$ & $0.846\pm0.098$ & $-0.197\pm0.084$ &  *****\\
47 Tuc   & $-0.70\pm0.10$ & $0.04\pm0.02$ & 1$1.0\pm1.4$ & $13.25\pm0.07$ & $0.690\pm0.096$ & $-0.149\pm0.081$ & $-1.280\pm0.072$\\
NGC 2420 & $-0.44\pm0.06$ & $0.05\pm0.02$ &  $2.0\pm0.3$ & $11.94\pm0.07$ & $0.498\pm0.118$ &  ***** & $-1.676\pm0.099$\\
\hline
\end{tabular}
\end{table*}

The two most important criteria in choosing open clusters for inclusion in this
work are that, firstly, each cluster must have a very well defined main 
sequence in the HR-diagram so that ridge lines can be derived and, secondly, 
the photometry (which is all existing in the literature) must go deep 
enough to apply our empirical MS-fitting method.  
This method uses only the unevolved lower portion of the MS in the fitting 
procedure, to ensure that there are no evolutionary effects -- hence the 
cluster photometry must extend down to at least $M_{V} = 7.0$.  Since we have 
$B$, $V$ and $I_{C}$ photometry for our 
field dwarf sample used in the MS fitting, we can utilize clusters with either
$(B-V)$ or $(V-I)$ photometry or, indeed, both.

We took, as a starting point, the sample of 8 open clusters from Sarajedini 
(1999, hereafter S99), however we discounted NGC~6791 and NGC~6819 as their 
main sequence photometry was not deep enough to apply our MS-fitting method 
(see Section 4.1 for more details).  
Thus we included M67, NGC~188, NGC~7789, Berkeley~39, 
NGC~2204 and Mel~66 -- the sources of photometry for these 6 clusters are 
as listed in S99, their Table 1.
We added 2 more young clusters for which there is deep photometry available 
in the literature, namely NGC~2204 (Kassis et al. 1997, $(B-V)$ and $(V-I)$ 
data) and NGC~2420 (Anthony-Twarog et al. 1990, $(B-V)$ data only).

Cluster abundances (and their quoted errors) were taken from the catalogue of 
Gratton (2000), ensuring that the metallicities for the sample clusters and 
for the field dwarfs used in the MS-fitting are on an homogeneous scale (see 
discussion in Paper I).  We took cluster reddenings from Twarog, Ashman \& 
Anthony-Twarog (1997), assuming a 0.02 mag error in $E(B-V)$ for all.
Absolute ages (and associated errors) were estimated from comparison of the 
dereddened cluster main lines, and their turn off absolute magnitudes, with 
Girardi et al. (2000) isochrones - the absolute magnitude of the turn off 
having been determined using the distance moduli derived in this paper.  
We note here that our derived ages do not differ significantly from those 
found by S99, who used the Bertelli et al. (1994) isochrones to estimate 
cluster ages.

These 8 open clusters span a metallicity range of 
$-0.44 \leq {\rm [Fe/H]} \leq +0.02$ and have ages from 1$-$7.5 Gyr.
We also included in the sample the metal-rich Galactic Globular Cluster 47~Tuc
(${\rm [Fe/H]} = -0.7$), since we have an empirically determined MS-fitting 
distance (and hence an age estimate) for this cluster which is also tied to 
the $Hipparcos$ distance scale (Percival et al. 2002).  
Photometry for 47~Tuc is from Stetson's `secondary' standards available 
through the Canadian Astronomy Data Centre web site 
(http://cadcwww.hia.nrc.ca/standards/ -- see Stetson 2000 for details).  
This extends the age range of the cluster sample up to 11 Gyr.
In addition, 6 of these clusters have $K$-band data from the Two Micron All 
Sky Survey (2MASS) Point Source Catalogue (http://irsa.ipac.caltech.edu/) 
which is used by Grocholski \& Sarajedini (2002) to determine the absolute 
$K$-band magnitude of the RC -- these are; NGC~2204, Berkeley~39, NGC~2420, 
NGC~2477, M67 and 47~Tuc.  The basic cluster parameters are listed in 
Table~\ref{tab_clus}.

\section{Method}
\label{method}

\subsection{Empirical MS fitting}
\label{msfit}

Before the MS-fitting method can be applied to derive the cluster distance,
the MS ridge line must be determined.  The procedure we adopted
consisted of making 0.2 mag cuts in $V$ across the MS and then plotting 
histograms in colour ($(B-V)$ or $(V-I)$) to find the peak value.  This method 
clearly identifies the ridge line for a cluster and also clearly shows the 
location of a binary sequence, if present.  The resultant points are fitted 
to a polynomial (usually cubic) function.

Cluster distances are derived using our empirical MS-fitting method, which we 
briefly outline here.  The method is fully described in Paper I, to which we 
refer the interested reader for more details. 
In Paper I we presented new $BVI$ (Johnson-Cousins) photoelectric photometry
for a sample of 54 field dwarfs with accurate $Hipparcos$ parallaxes, which 
span a metallicity range of $-0.4 \leq {\rm [Fe/H]} \leq +0.3$.  Absolute
magnitudes for these stars are in the range $5.4 \leq M_{V} \leq 7.0$, 
thus they are unevolved zero-age MS stars and their location in a 
Colour-Magnitude diagram is totally insensitive to any age differences between
the clusters which we are fitting to and field dwarf sample. 

The basic MS-fitting method consists of constructing a template MS from the
field dwarf sample by applying colour shifts to the individual stars to 
account for the differences in metallicity between the field stars and the 
cluster.  This template is then shifted in $V$ to match the dereddened and
extinction-corrected cluster main line, the difference in magnitude being 
equal to the absolute distance modulus $(m-M)_{0}$.  

The procedure used to calculate the magnitude of the metallicity dependent 
colour shifts for the field dwarfs relies on first establishing
that the shape of the MS is insensitive to [Fe/H] in the narrow range of 
metallicities and magnitudes we are dealing with.  Next, we determine the 
colour that each field dwarf would have at a fixed magnitude of $M_{V}=6$, 
which we call $(B-V)_{M_{V=6}}$ (or $(V-I)_{M_{V=6}}$), using the slope of 
the Hyades MS as a reference slope.  
Finally, we calculate the derivative $\delta(B-V)_{M_{V=6}} / \delta[Fe/H]$
(or $\delta(V-I)_{M_{V=6}} / \delta[Fe/H]$).
Because of the constant shape of the MS across the metallicity range, this 
derivative is appropriate for the whole magnitude range spanned by the field 
dwarfs, and hence colour shifts are applied to each star in the sample, at 
their observed magnitudes, to construct a template MS at the metallicity of 
the cluster in question.

\begin{figure}
\psfig{file=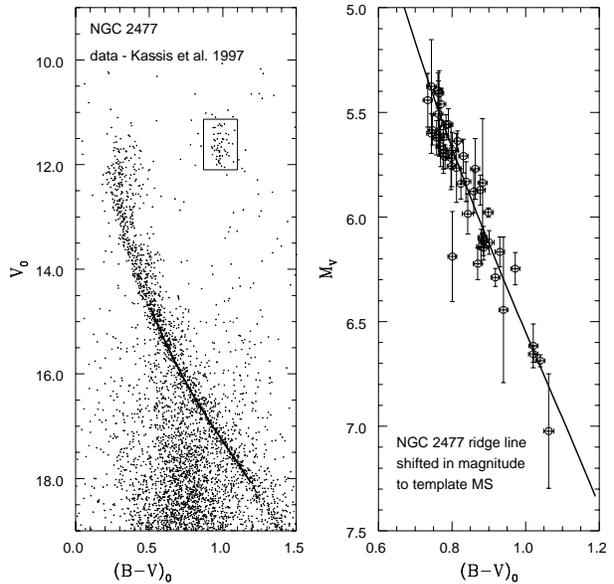,width=8.3cm}
\caption{Illustration of MS-fitting method -- see text for details.}
\label{fit}
\end{figure} 

When fitting to the cluster main lines to derive the cluster distance, weighted
errors are used for the field dwarfs which include both photometry errors
and magnitude errors due to error on the parallax (where 
$\Delta M_{V} = 2.17(\Delta\pi/\pi)$).  We also account for errors on the 
metallicity of the both the field dwarfs and the cluster, and the best-fitting 
distance is found by minimizing $\chi^{2}$.
For clusters with photometry in both colour indices, dereddened distances are
taken to be an average from the $(B-V)$ and $(V-I)$ fits, which we note here
are always consistent with each other, within their respective errors.

The fitting procedure is illustrated in Figure~\ref{fit} which shows the 
$V/(B-V)$ CMD for NGC~2477 (data from Kassis et al. 1997).  The left panel
shows the dereddened and extinction corrected cluster photometry overlaid with
the derived ridge line for the MS.  The RC stars are contained within the box
spanning $11.1 \leq V_{0} \leq 12.1$ and $0.85 \leq (B-V)_{0} \leq 1.1$.  The
panel on the right shows the cluster ridge line, shifted in $V$-magnitude
to match the template MS created from the field dwarf sample, hence yielding 
the best-fitting distance modulus.  The data points are the template MS, which
has been created from the field dwarf sample by `correcting' their colours to 
match the metallicity  of the cluster (${\rm [Fe/H]} = 0.0$ in this case), as
described above.  The dereddened distance moduli determined by this method for
all the clusters in the sample are listed in Table~\ref{tab_clus}.

\subsection{RC magnitudes - observed}
\label{rcobs}

Apparent RC magnitudes, $m_{\lambda}^{RC}$, are taken as listed in S99 
($V$- and $I$-bands only) for M67, NGC~188, NGC~7789, Berkeley~39, NGC~2204 
and Mel~66.  For NGC~2204 and NGC~2420 we calculated the RC magnitude using 
exactly the same method employed by S99, i.e. finding the median magnitude of 
the stars contained in a box centered on the RC.  As noted by S99, this median
value is virtually insensitive to the exact location of the box, a fact 
confirmed by us from Monte Carlo simulations of the RC (see Sect.~\ref{rcthe}).
$K$-band RC magnitudes, determined using the same method, are taken from 
Grocholski \& Sarajedini 2002 (note that these are listed as extinction
corrected absolute magnitudes, $M^{RC}_{K}$, so we have used their 
quoted reddening and distance modulus values to calculate the apparent $K$ 
magnitudes, $m_{K}^{RC}$).
 
Thus, for each cluster the absolute RC magnitude, $M_{\lambda}^{RC}$ is given 
by:
\begin{equation}
M_{\lambda}^{RC}(cluster) = m_\lambda^{RC} - A_{\lambda} - (m-M)_{0}
\label{rc_eq}
\end{equation}
where $m_\lambda^{RC}$ is the apparent RC magnitude as described above, 
$A_{\lambda}$ is the extinction and $(m-M)_{0}$ is the dereddened distance
modulus as found in Sect.~\ref{msfit}.
 
The population correction, $\Delta M_{\lambda}^{RC}$, is then given by
\begin{equation}
\Delta M_{\lambda}^{RC} = M_{\lambda}^{RC}(local) - M_{\lambda}^{RC}(cluster)
\label{deltam_eq}
\end{equation}
where $M_{\lambda}^{RC}(local)$ are the values given by Alves (2002), for $V$, 
$I$ and $K$ (see Sect.~\ref{RCasDI}).

\subsection{RC magnitudes - theoretical}
\label{rcthe} 

Mean RC magnitudes generated from the Girardi et al. (2000) models are
tabulated in GS01 ($V$ and $I$) and SG02 ($K$) for a range of ages and
metallicities (see Sect.~\ref{intro}).  These effectively correspond to 
cluster RC magnitudes, $M_{\lambda}^{RC} (cluster)$, since each tabulated 
value represents the RC magnitude for a single age and metallicity.

For the mean magnitude of the local RC, $M_{\lambda}^{RC}(local)$, we use the 
values calculated by GS01 and SG02.  GS01 and SG02 employ a complete population
synthesis algorithm which, using the stellar models of Girardi et al. (2000), 
produces a synthetic CMD, from which the RC luminosity function can be 
extracted.  The solar neighbourhood was modelled using the Star Formation Rate
(SFR) and Age Metallicity Relation (AMR) from Rocha-Pinto et al. (2000a,b).
We will discuss the effects of the particular choice of SFR and AMR in 
Section~\ref{disc}.  The model population correction is then calculated as for 
the empirical data i.e. $\Delta M_{\lambda}^{RC} = M_{\lambda}^{RC}(local) - M_{\lambda}^{RC} (cluster)$.  It is important to note at this point that we are 
not relying on the \emph{absolute} values for the RC magnitudes from the models
since the population correction is a differential quantity, i.e. it is the
\emph{difference} between the local RC magnitude and the RC magnitude of the 
population under scrutiny.
 
The tabulated model values give the mean properties of the RC stars -- i.e.
the mean magnitude of all stars in the simulation which are still undergoing 
core helium burning.  RC stars start their lives on the ZAHB, producing a peak
in their luminosity function, and then evolve on shorter timescales to 
brighter magnitudes.
Consequently, the mode of the magnitude distribution is offset from the mean 
value, in the sense that the mean magnitude is brighter.  Obviously, from the 
simulations, all the RC stars are accounted for and any value can be 
measured -- mean, median or mode.
For real data sets the RC is often sparsely populated, due to poor sampling, 
and the empirical median values derived as described in Sect.~\ref{rcobs} 
reflect the mode rather than the mean magnitude.  
  
We tested for the offset between mean and median magnitudes by first 
simulating RCs at ages from 1--10 Gyr in $V$, $I$ and $K$.  This was done by 
populating the Girardi et al. (2000) isochrones for solar metallicity using a
Salpeter IMF, and then taking all stars still in their core helium-burning 
phase.  These stars are all within 1.0 mag of the ZAHB and, in the particular 
simulations used, we generate a total parent sample of 4000--5000 RC stars.
(We note here that the particular choice of IMF does not influence the results
at all, since the evolutionary time-scale for these stars is very short, and 
so the mass is nearly constant in the region of interest).
We then randomly extracted subsamples of 5 and 50 stars from the parent sample 
and calculated the median magnitude in all 3 pass-bands for each subsample. 
This process was repeated 70 times and the resultant median magnitudes plotted
as histograms.  The peak of each distribution (i.e. for the subsamples of 
5 and the subsamples of 50) yields the most likely measured median value for
a real data set, and 
the dispersion (i.e. 1-$\sigma$) of the distribution represents the error 
associated with the (small) sample size.  The median values were found to be 
identical (within 0.01 mag) for the simulated subsamples of 5 and 50, and 
whilst the errors are obviously larger for the smaller sample size, they are 
surprisingly stable -- 
errors are generally no more than 0.05 mag for a sample of 5, except for the 
1~Gyr simulation which produces a 1-$\sigma$ error of $\sim$ 0.1 mag.  
However, this is not a cause for concern here, as all the young clusters 
used in this study have a well populated RC, which produces much smaller 
errors.

The mean--median offsets are 0.04--0.05 mag for ages $>$ 2 Gyr in all 3 
filters, whilst at 1 Gyr, the offset is 0.07--0.08 mag.
Our empirical RC absolute magnitudes, $M^{RC}_{\lambda}$, were therefore 
adjusted using the appropriate mean$-$median offset before calculating the 
population corrections, $\Delta_{\lambda}^{RC}$, and comparing with the 
theoretical values.  
The empirical values of $M^{RC}_{\lambda}$ are listed in 
Table~\ref{tab_clus} -- the quoted errors include the error on the cluster 
distance modulus, the statistical error on the measured RC magnitude, 
described above, and the error induced by a 0.02 mag uncertainty on the 
cluster reddening, added in quadrature.

\section{Results}
\label{results}

\begin{figure}
\psfig{file=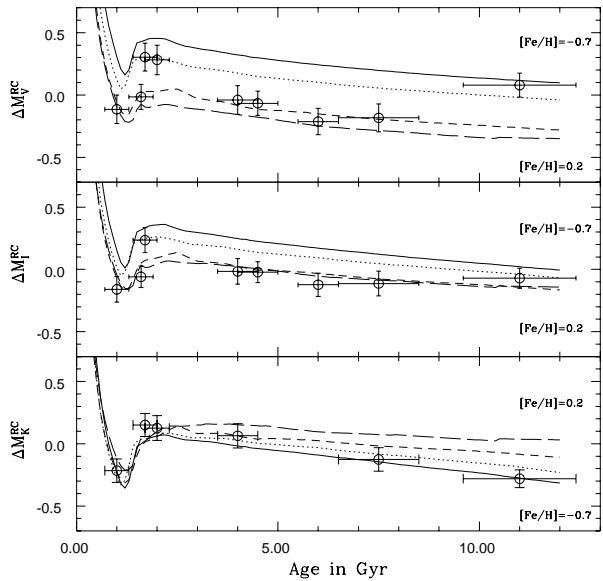,width=8.3cm}
\caption{The RC population correction, $\Delta M^{RC}_{\lambda}$, in $V$, $I$ 
and $K$, plotted against age.  
The lines are the Girardi et al. (2000) models at [Fe/H] 
of; -0.7 (solid), -0.4 (dotted), 0.0 (short--dash), +0.2 (long--dash). 
The data points are the empirically derived values for 9 (in $V$), 8 (in $I$)
and 6 (in $K$) clusters, with error bars as detailed in the text.}
\label{dmrc}
\end{figure} 

Fig.~\ref{dmrc} shows the empirically derived population corrections, 
$\Delta M_{\lambda}^{RC}$, in $V$ (9 pts), $I$ (8 pts) and $K$ (6 pts) 
overplotted with the theoretical models of Girardi et al. (2000) for [Fe/H] 
between $-$0.7 and $+$0.2.  The data points are our empirically determined 
values and the vertical error bars include the error on our distance moduli 
(which includes error on cluster metallicity), the 
statistical error on the RC magnitude (detailed in Sect.~\ref{rcthe}), and the 
error on the RC magnitude induced by uncertainty in the cluster reddening 
(0.02 mag), all added in quadrature.
The most important feature of this plot is the fact that the empirically 
determined population corrections follow the trend of the theoretical models
in all 3 pass-bands, particularly at ages $<$ 2 Gyr, where the models
predict a strong dependence on age. 

\begin{figure}
\psfig{file=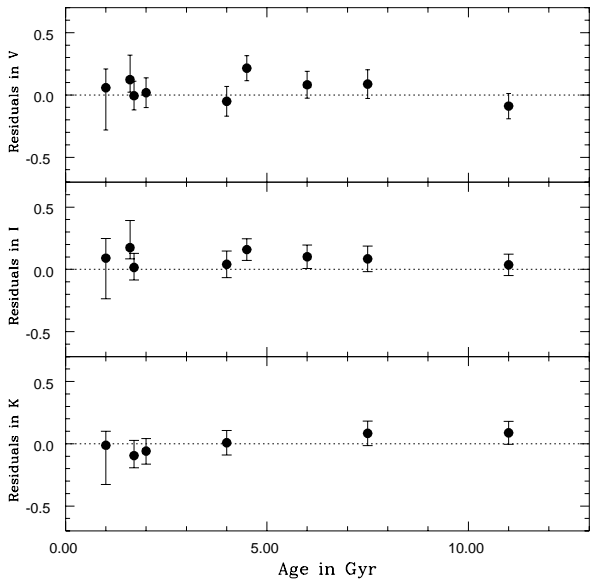,width=8.3cm}
\caption{The residuals in the population correction, $\Delta M^{RC}_{\lambda}$,
in the sense $model - observed$, plotted against age.}
\label{res_age}
\end{figure} 
\begin{figure}
\psfig{file=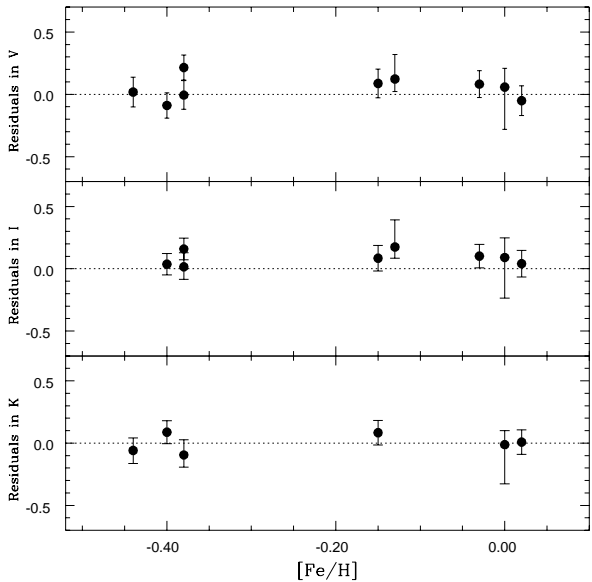,width=8.3cm}
\caption{Residuals in $\Delta M^{RC}_{\lambda}$ as for Fig.~\ref{res_feh},
plotted against [Fe/H].}
\label{res_feh}
\end{figure} 
We stress again that we are not relying on 
\emph{absolute} values from the models, which may differ from the empirical 
ones (e.g. for local RC magnitude, see GS01).  The models are always being
used differentially, remembering that the population correction,
$\Delta M_{\lambda}^{RC}$, is the \emph{difference} between the local RC 
magnitude and the RC magnitude of the cluster in question.

In order to quantify the goodness-of-fit between the models and the empirical
data, we calculated the residuals of $\Delta M_{\lambda}^{RC}$ in the sense 
$model - observed$.  These residuals are plotted in Figs.~\ref{res_age} and 
\ref{res_feh} against, respectively, age and metallicity.  The vertical 
error bars (i.e. in magnitude only) are as in Fig.~\ref{dmrc}, but have been 
amended to account for the error on the cluster age.  It can be seen from 
Fig.~\ref{dmrc} that the error on the cluster age will cause an asymmetric 
error on $\Delta M^{RC}_{\lambda}$, particularly for ages $<$ 2 Gyr, and this 
is reflected in the error bars in Figs.~\ref{res_age} and \ref{res_feh}.  
These plots clearly show that the residuals in $\Delta M^{RC}_{\lambda}$ 
display no statistically significant trends with either age or metallicity 
in any of the 3 pass-bands.

The most discrepant point in the $V$- and $I$-bands (i.e. with the largest 
residuals) is Melotte 66.  This is the only cluster for which there is no
$(B-V)$ data for the main sequence and so the distance modulus has been derived
from the $(V-I)$ data alone.  We note, however, that there is a small amount 
of photoelectric $UBVRI$ photometry for this cluster from Twarog, 
Anthony-Twarog \& Hawarden (1995).  Comparison of the photoelectric data
with the CCD data used in this study (from Kassis et al. 1997) indicates that 
there may be small zero-point offsets in the CCD photometry which would have
the effect of making the $(V-I)$ colours too red by 0.04--0.05 mag.  If this
were the case, the derived distance would be too short by $\sim$ 0.2 mag, 
causing a similar discrepancy in the RC magnitude, and therefore accounting
for the discrepancy between the model and observed values for the population
correction.  Removing this cluster from our analysis, the mean residuals 
in $\Delta M^{RC}_{\lambda}$ are; +0.03 $\pm$0.07 in $V$, +0.07 $\pm$0.06 
in $I$, and 0.00 $\pm$0.07 in $K$, where the quoted errors are the 1-$\sigma$ 
errors.  Hence we find no statistically significant offsets in $V$ and $K$
and a very marginal offset in $I$, significant only at the 1.2-$\sigma$ level.
 
We note that if we included Melotte 66 in this analysis the mean residuals 
in $V$ and $I$ would become +0.05 $\pm$0.09 and +0.09 $\pm$0.06 respectively
(those in $K$ would remain unchanged) and hence our conclusions would 
essentially remain unaltered.

\section{Discussion and conclusions}
\label{disc}

A vital factor in determining the RC population corrections is the absolute
magnitude of the local RC, for which we have very well determined 
observational values.  Deriving the same quantities from the models requires
knowledge of the local SFR and AMR, which are used as input parameters in the
population synthesis algorithm (see GS01 for details).  We checked the 
effects of assuming a different SFR and/or AMR by modelling the local RC 
using the AMRs of Carlberg et al. (1985) and of Edvardsson et al. (1993).  We 
also tested the effect of using a constant star formation rate.  In all cases,
the model values for the mean RC magnitude, $M^{RC}_{\lambda}$, were 
indistinguishable from those found when assuming the Rocha-Pinto et al. 
(2000a,b) relationships.  This is largely because the local RC is dominated by
stars with ages $<$ 2 Gyr (see in depth discussion in GS01) and the
different AMRs all predict similar metallicities at these young ages.  Hence 
the precise details of the star formation history do not significantly affect
the local RC magnitude.

The local RC simulated in GS01 (using the Rocha-Pinto 2000 SFR and AMR) 
yields a straight mean metallicity of ${\rm [Fe/H]} = -0.04$, 
whilst the observed RC stars in the solar neighbourhood appear to have a mean 
[Fe/H] of $\sim$ -0.18 (see, e.g. Alves 2000).  Metallicities used by Alves
for the local RC stars are from McWilliam (1990) -- the same ones used by 
Udalski (2000) to derive a metallicity dependence for $M^{RC}_{I}$.  This 
offset between the model and observed mean metallicity for the RC stars has
already been noticed and discussed in GS01.  
McWilliam himself notes that his metallicities for the RC stars may 
be systematically too low, especially for the highest metallicities, due to 
the (then) lack of suitable model atmospheres for super-solar metallicities.  
Luck \& Challener (1995) determined abundances for 55 G and K field giants
from high resolution spectra -- 45 of these stars being in common with the 
McWilliam (1990) sample.  Luck \& Challener find a constant offset of 0.12 dex 
between their abundances and those of McWilliam, in the sense that the 
McWilliam (1990) values are lower.  From the most recent catalogue of [Fe/H] 
determinations for FGK stars by Cayrel de Strobel, Soubiran \& Ralite (2001),
we made a comparison of all the stars in the McWilliam (1990) sample which had 
abundance determinations by other authors -- this yielded a sample of 89 stars
with 152 measurements from different sources.  The average difference between
the McWilliam abundances and all the others in the literature is 0.13 dex, with
a dispersion of 0.15 dex.  However it is evident that the McWilliam 
metallicities are increasingly discrepant at the highest metallicities.  
Splitting the sample into 2 groups with ${\rm [Fe/H]}_{\rm other} < 0.2$ and 
${\rm [Fe/H]}_{\rm other} > 0.2$ and comparing with the McWilliam values
reveals an offset of 0.10 dex for the lower metallicity group and an offset
of 0.31 dex for the higher metallicity group, both with a dispersion of 0.14 
dex, in the sense that the McWilliam values are lower.  Hence, the apparent
discrepancy between the average metallicity of the local RC predicted from the 
models and that derived from observational data, appears to be a problem with 
the zero-point of the metallicity scale derived by McWilliam (1990).  


The results of our analysis fully confirm the theoretical population 
corrections obtained by GS01 and SG02 for single-age, single-metallicity 
stellar populations, which are the building blocks for computing corrections 
to composite systems.
Remarkably, the minimum of the RC luminosity for ages around 1 Gyr predicted 
by theory, corresponding to the transition from electron degenerate to 
non-degenerate helium cores, is also clearly observed. This is a further 
positive test for the accuracy of the physics employed in the stellar models.
A larger sample of clusters with accurate multicolour photometry for the
lower MS could confirm whether the small offset in the $I$-band is real.
For use as an accurate distance indicator, $K$ is the best single pass-band 
to use as the effects of reddening are minimised and, provided 
ages are less than $\sim$ 7 Gyr, the metallicity dependence is very small.

It is evident from Fig.~\ref{dmrc} that, in general, there is no 
simple relationship between 
the population correction in a given passband and metallicity or age. Also,
it is only in the $K$-band, and specifically for metallicities 
close to solar and ages lower than a few Gyr, that the population corrections 
are very small (see SG02 for a discussion on this point).
This means that there is no way, a priori, to determine generic empirical 
population corrections that are a function of only the metallicity of the 
population under scrutiny (this point has already been discussed in depth by 
GS01). In the case of composite stellar populations it is necessary to have
a good determination of their Star Formation Rate (SFR) and Age 
Metallicity Relationship (AMR) and apply population synthesis algorithms 
in order to obtain the appropriate corrections (alternatively, the 
corrections can be derived using the simpler analytical method described 
in GS01). The accuracy of 
the derived corrections will depend entirely on the accuracy of the determined
SFR and AMR, since they are combinations of the single-age, single-metallicity
corrections, which we have now proved to be accurate.
For systems where no precise AMR and SFR are available, the use of empirical 
calibrations --  which are necessarily of limited validity since they are 
dependent on the particular properties of the calibrating sample -- can be
very dangerous because it can lead to errors of several tenths of a magnitude 
on the derived distance moduli.

The very limited range of validity of empirical corrections based only on 
the stellar metallicity is illustrated by the valuable data published by 
Pietrzy\'nski, Gieren \& Udalski~(2003, hereafter PGU03). These authors 
present new near-infrared $J$- and $K_{\rm S}$-band data for several fields in
the LMC, SMC, and the Fornax and Carina dwarf galaxies and, in addition, they 
utilize $I$-band data in the literature for the same systems.  PGU03 are 
therefore able to determine accurately the $K$- and $I$-band brightness of 
the RC in these 4 systems and they go on to derive empirical corrections to 
the RC brightness, based solely on differences in mean metallicity.
After applying these corrections (which are normalized to the LMC mean 
metallicity) to the RCs in the SMC, Carina, Fornax and the solar 
neighbourhood, 
PGU03 look at the difference between the $I$ and $K$ brightness of the RC for 
each system.  If the corrections are adequate, one expects that this quantity
will be constant for all the systems, however PGU03 find a discrepancy of 
0.23 mag for the solar neighbourhood value, which they ascribe to zero-point 
errors in the photometry of the local RC stars.

This difference between the $I$- and $K$-band absolute magnitude of the RC 
is an ideal parameter on which we can test the theoretical models further. 
The predicted metallicity dependencies for the population correction, 
$\Delta M^{RC}_{\lambda}$, are very different between the 2 passbands, that in 
$I$ being much stronger than in $K$.  Furthermore, the dependencies work in 
opposite directions -- for increasing metallicity, $\Delta M^{RC}_{I}$ becomes
smaller, whilst $\Delta M^{RC}_{K}$ becomes larger -- and are not constant
with increasing age.  
Therefore, as an experiment, we examined the difference between the $I$-band 
and $K$-band magnitude of the RC predicted from theory for the LMC, SMC, 
Carina and the solar neighbourhood.  It is important to
realise that in case of the theoretical population corrections, consistency 
with the local RC is automatically achieved, since the corrections are 
computed relative to the solar neighbourhood values.  We have adopted the 
values for the LMC and SMC provided by SG02; for Carina, we have simulated
the RC using the same population synthesis algorithm, assuming the SFR used
in GS01 and adopting the most recent determination of the AMR from 
Tolstoy et al. (2003).  We did not consider the case of Fornax simply because 
existing determinations of the SFR are highly uncertain, 
thus preventing an accurate assessment of the appropriate theoretical 
population corrections.

\begin{figure}
\psfig{file=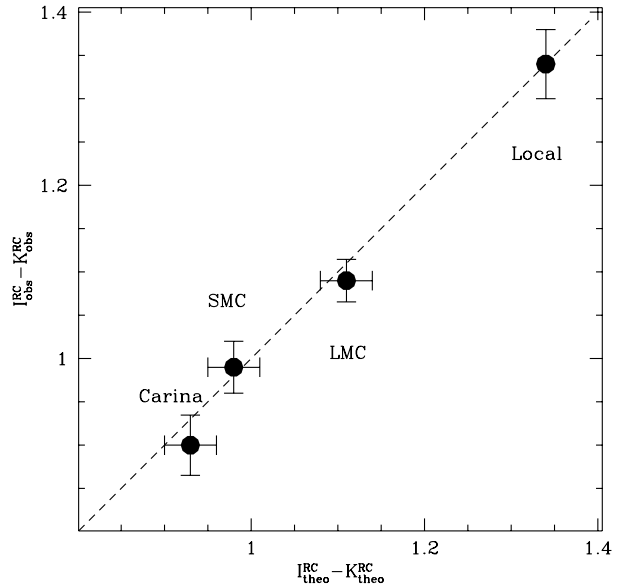,width=8.3cm}
\caption{The difference between the $I$-band and $K$-band magnitude of the Red 
Clump for the LMC, SMC, Carina and the solar neighbourhood -- observed 
values vs. theoretical predictions.  To compute the vertical error bars we 
have summed in quadrature the 
error on the photometric determinations of the RC in $K$ and $I$ as given 
by PGU03, and the contribution due to the error on the adopted reddening.
The horizontal error bar reflects the error on the determination of the 
local RC magnitude, which is added to the theoretical population correction.
The dashed line represents a one-to-one correspondence, and is not a fit to 
the points.}
\label{lmc_smc}
\end{figure} 

Figure~\ref{lmc_smc} shows a comparison of the observed difference
between the $I$- and $K$-band brightness of the RC, for the LMC, SMC, Carina
and the solar neighbourhood, and the corresponding values predicted from 
theory.
(We note here that the $K_{\rm S}$-band magnitudes from PGU03 have been 
transformed to the Bessell \& Brett system, which are those used in the 
theoretical models).
The result of the comparison is excellent, with a good agreement, within the 
errors, between theory and observations.  This confirms the accuracy of the 
theoretical corrections, which simultaneously reproduce the LMC, SMC and Carina
data and satisfy the solar neighbourhood values without the need to invoke 
photometric errors of the order of 0.20 mag. As a consequence, 
with the theoretical population corrections one obtains full consistency 
between the RC $I$- and $K$-band distances to the LMC derived from the 
empirical values tabulated by PGU03;
in fact, by using the local RC calibration by Alves et al.~(2002) we obtain, 
respectively, $(m-M)_{0, I}$=18.49$\pm$0.06 and $(m-M)_{0, K}$=18.46$\pm$0.03
(the errors include both the error on the reddening and the error on the 
local RC magnitudes).
For the SMC, again using the observed values tabulated by PGU03, we obtain  
$(m-M)_{0, I}$=18.90$\pm$0.06 and $(m-M)_{0, K}$=18.88$\pm$0.03.
The difference in the RC distance moduli, $\Delta M_{LMC-SMC}$, is equal to
0.41$\pm$0.08 in $I$ and 0.42$\pm$0.04 in $K$, which compares well
with the value of 0.50$\pm$0.08 obtained from the comparison of the Tip of the
Red Giant Branch magnitudes found by PGU03. 

Significantly, the LMC distance moduli we find above, derived individually 
from the $I$- and $K$-band data, are fully in agreement with the LMC distance 
modulus determined from the multiwavelength method of Alves et al. (2002).
This method imposes the constraint that the derived distance modulus must be
the same in all the passbands used  -- solving simultaneously in several 
passbands then yields both distance modulus and reddening.
Using $V$, $I$ and $K$ data, Alves et al. (2002) find 
$(m-M)_{0, LMC} = 18.493 \pm0.033_{\rm r} \pm0.03_{\rm s}$.  
Most recently, Salaris et al. (2003) have applied the same method to the field 
surrounding the LMC cluster NGC 1866 and found 
$(m-M)_{0, LMC} = 18.53 \pm0.07$ from $HST$ $V$ and $I$ data.

\section*{Acknowledgments}

This research is based on observations made at the South~African~Astronomical
Observatory.

We have made extensive use of the WEBDA database 
(http://obswww.unige.ch/webda).  
We thank the anonymous referee for a prompt and careful reading of the 
manuscript.
SMP acknowledges financial support from PPARC.


\label{lastpage}

\end{document}